\documentclass[osajnl2,twocolumn,showpacs,superscriptaddress]{revtex4}

 \usepackage{amsmath}

 \usepackage{graphicx}
 \graphicspath{{pictBM/}{pict/}{}}
 \usepackage{amssymb}
 \usepackage{bm}
 \usepackage{here}
 
\newcommand{\ud}{\mathrm{d}}

 \newcommand\pictc[5]{\begin{figure}
             \centerline{\vspace{0mm}
 \includegraphics*[width=\columnwidth,height=\textheight,keepaspectratio]{#3}}
             \protect\caption{\protect\label{fig:#4} #5}
                     \end{figure}            }

 \newcommand\pict[4][1]{\pictc{#1}{!htb}{#2}{#3}{#4}}
 \newcommand\rpict[1]{\ref{fig:#1}}

 \newcommand\leqt[1]{\protect\label{eq:#1}}
 \newcommand\reqtn[1]{\ref{eq:#1}}
 \newcommand\reqt[1]{(\reqtn{#1})}

 \newcounter{Fig}
 
\begin{document}

\title{Bloch-mode extraction from near-field data in periodic waveguides}

\author{Sangwoo Ha}
\author{Andrey A. Sukhorukov}
\affiliation{Centre for Ultra-high bandwidth Devices for Optical Systems (CUDOS), Nonlinear Physics Centre, Research School of Physics and Engineering, Australian National University, Canberra, ACT 0200, Australia}

\author{Kokou B. Dossou}
\author{Lindsay C. Botten}
\affiliation{CUDOS, Department of Mathematical Sciences, University of Technology, Sydney,
New South Wales 2007, Australia}

\author{C. Martijn de Sterke}
\affiliation{CUDOS, School of Physics, University of Sydney, New South Wales 2006, Australia}

\author{Yuri S. Kivshar}
\affiliation{Centre for Ultra-high bandwidth Devices for Optical Systems (CUDOS), Nonlinear Physics Centre, Research School of Physics and Engineering, Australian National University, Canberra, ACT 0200, Australia}

\begin{abstract}
We demonstrate that the spatial profiles of both propagating and evanescent Bloch-modes in a periodic structure can be extracted from a single measurement of electric field at the specified optical wavelength. We develop a systematic extraction procedure by extending the concepts of high-resolution spectral methods previously developed for temporal data series to take into account the symmetry properties of Bloch-modes simultaneously at all spatial locations.
We demonstrate the application of our method to a photonic crystal waveguide interface and confirm its robustness in the presence of noise.
\end{abstract}

\ocis{\small (050.5298) Photonic crystals;
                    (230.7370) Waveguides;
                    (250.5300) Photonic integrated circuits.}

\maketitle

Periodically modulated optical waveguides offer new possibilities for controlling the propagation of light. Resonant scattering from periodic modulations can be used to tailor the dispersion, enabling in particular a dramatic modification of the group velocity and realization of slow-light propagation. Such fundamental effects can be directly visualized in experiment with near-field measurements, which can be used to recover the amplitude, phase, and polarization of the electric field at all spatial locations in the plane of the waveguide~\cite{Engelen:2007-401:NAPH}. This information can then be used to extract the dispersion characteristics of the guided modes.

A commonly used approach to the dispersion extraction is through the spatial Fourier-transform (SFT) of the field profiles, since peaks in the Fourier spectra correspond to the wavenumbers of guided modes~\cite{Gersen:2005-123901:PRL, Gersen:2005-73903:PRL, LeThomas:2008-125301:PRB}. However, there exists {a fundamental limitation} on results obtained with SFT: $\Delta k \ge 2 \pi / L$, where $\Delta k$ is the resolution of the wavenumber, and $L$ is the structure's length. Therefore, accurate dispersion results can only be obtained for long waveguides, extending over many periods of the underlying photonic structure.
Another limitation of the SFT method is that it cannot provide information on the dispersion of evanescent waves, which may play an important role close to the structure boundaries or interfaces between different waveguides.
For example, evanescent waves enable efficient excitation of slow-light waves without a transition region~\cite{White:2008-2644:OL}.

Alternative methods for dispersion extraction have been developed to overcome the shortcoming of the SFT method.
It was shown that an interference of two counter-propagating modes can be used to extract their wavenumbers~\cite{Fan:1999-3461:APL}, however this technique is not applicable under the presence of multiple propagating modes or evanescent waves.
Recently, it was demonstrated that dispersion extraction in multi-mode waveguides with in principle unbounded resolution is possible even for short waveguide sections~\cite{Dastmalchi:2007-2915:OL, Sukhorukov:2009-3716:OE}, using approaches based on an adaptation of high-resolution spectral methods previously developed for the analysis of temporal dynamics~\cite{Roy:1991-109:PRP, Mandelshtam:2001-159:RAR}. In this work, we introduce an important generalization of such methods taking into account the spatial symmetry properties of modes in periodic waveguides. We show that beyond the dispersion relations, it is possible to {\em extract the spatial profiles of all guided modes}. Our method is applicable to an arbitrary combination of propagating and evanescent waves. We illustrate the application of this general approach by analysing light dynamics at an interface between photonic crystal waveguides designed for coupling into a slow-light mode~\cite{White:2008-2644:OL}

\pict{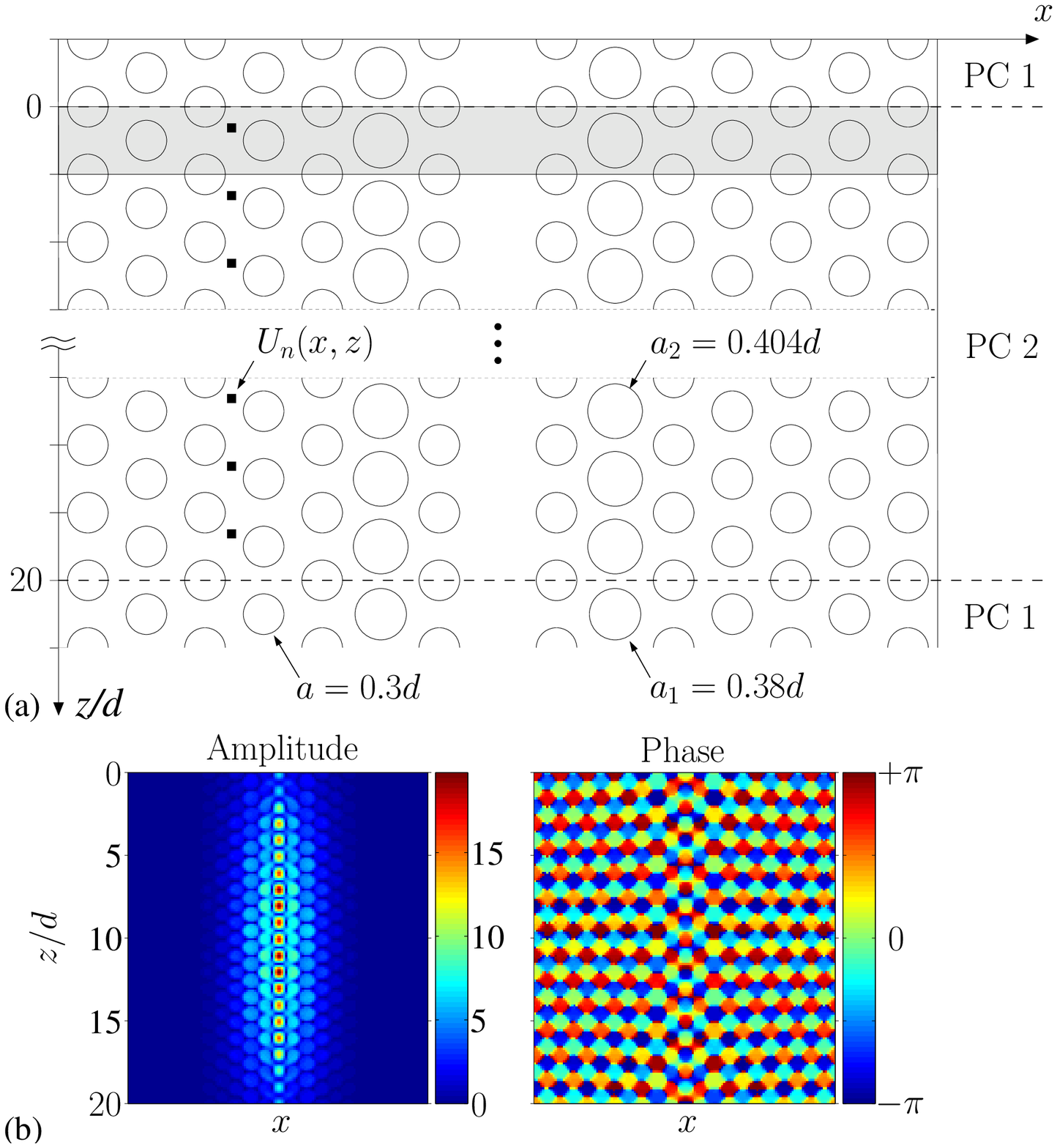}{schematic}{
(a)~Schematic of a two-dimensional photonic crystal waveguide. The Bloch-wave extraction is performed in the PC2 section.
(b)~Numerically calculated amplitude (left) and phase (right) of complex electric field profile at the normalized frequency $d/\lambda \simeq 0.2662$.
}

Let us consider a periodic waveguide section, where the light propagation in a particular frequency range is primarily determined by a finite total number of guided modes ($M$). The value of $M$ can be established based on numerical modelling, taking into account both propagating and evanescent waves.
Since each of the modes of a periodic waveguide satisfies the Bloch theorem~\cite{Joannopoulos:1995:PhotonicCrystals}, the complex electric field envelope of a waveguide mode with the index $m$ at the frequency $\omega$ can be expressed as
$\psi_m(\mathbf{r}; \omega) \exp( i k_m z / d )$.
Here $k_m$ are the complex Bloch wavenumbers, $\mathbf{r}=(x,y,z)$ where $x$ and $y$ are the orthogonal directions transverse to the waveguide and $z$ is the direction of periodicity, $d$ is the waveguide period, and $\psi_m$ are the periodic Bloch-wave envelope functions: $\psi_m(z) = \psi_m(z+d)$.
Then, the total field inside the waveguide can be presented as a linear superposition of $M$ propagating modes with amplitudes $a_m$ and small additional contributions $w(\mathbf{r};\omega)$:
\begin{equation} \leqt{E}
   E(\mathbf{r}; \omega) = \sum_{m=1}^M a_m \psi_m(\mathbf{r}; \omega)\exp( i k_m z / d ) + w(\mathbf{r};\omega).
\end{equation}
Here $w(\mathbf{r};\omega)$ can account for the radiation field due to the excitation of non-guided waves and for vanishingly small evanescent waves which are excluded from consideration, and this term can also appear due to noise in experimental measurements.

\pict{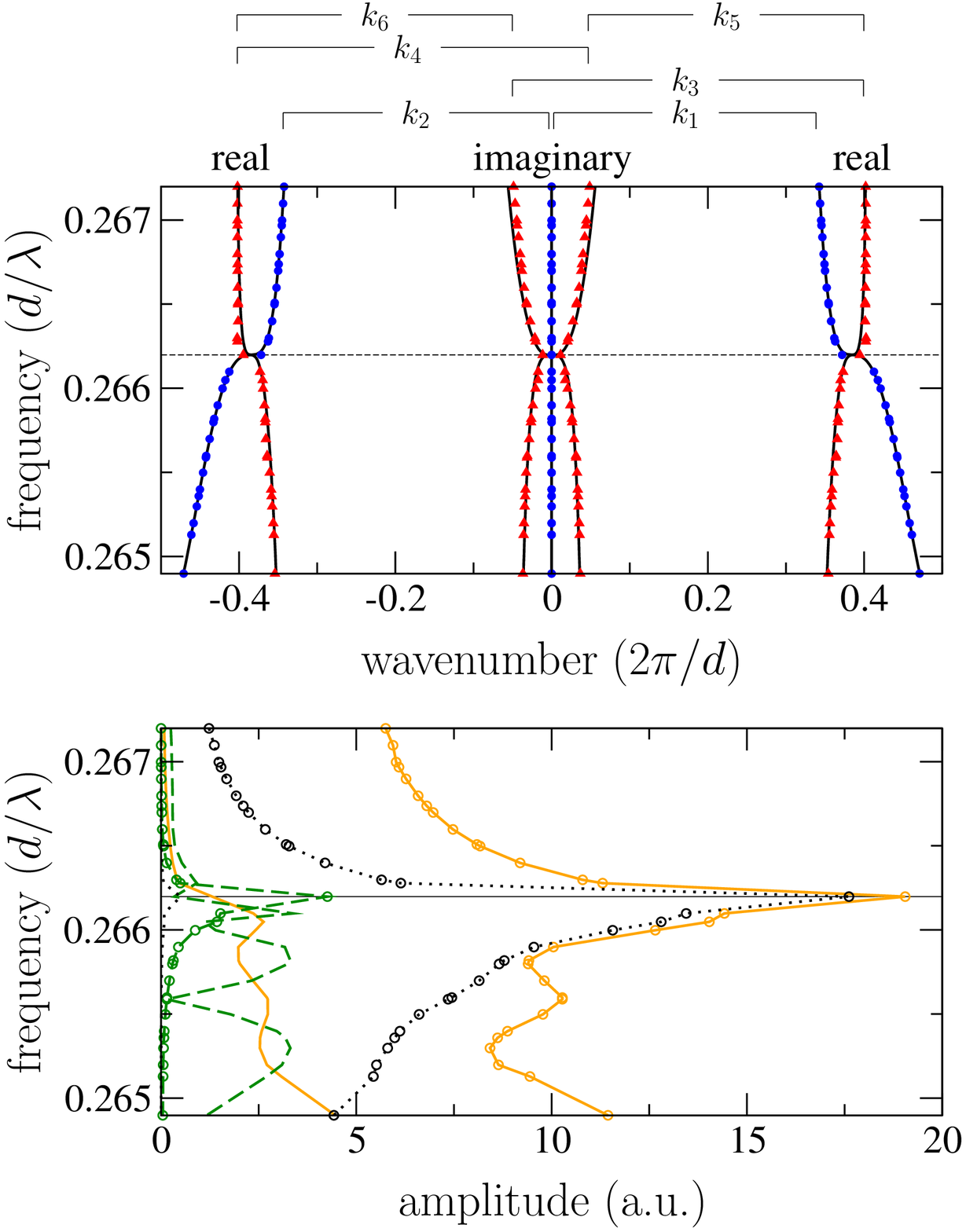}{wavenumber}{
Complex wavenumbers $k_m$ extracted with spatial spectral analysis. Propagating and evanescent modes are marked by circles and triangles, respectively.
Solid curves show numerically calculated dispersion. Real and imaginary parts of each mode are indicated above.
Horizontal lines at $d/\lambda\simeq0.2662$ mark the frequency at the inflection point.}

We now present the procedure for the simultaneous extraction of the wavenumbers of profiles of the guided modes. Let us separate the spatial domain in a number of unit cells, $(x,y,z+n\cdot d)$. Here $d$ is the period in the direction along the waveguide, $z_0$ belongs to a single unit cell ($z_{\rm min}<z<z_{\rm min}+d$), and $n=1:N$ where $N$ is the number of cells in the waveguide section. Let us denote with $U_n(\mathbf{r};\omega) = E(x,y,z+n\cdot d; \omega)$, $A_m(\mathbf{r};\omega) = a_m \psi_m(\mathbf{r};\omega)$, and
$w_n(\mathbf{r};\omega) = w(\mathbf{r}+n\cdot d;\omega)$, where $\mathbf{r}$ belongs to the first unit cell.
Then Eq.~\reqt{E} can be written as,
\begin{equation} \leqt{U}
  U_n(\mathbf{r};\omega) = \sum_{m=1}^M A_m(\mathbf{r}; \omega) \exp( i k_m n ) + w_n(\mathbf{r};\omega),
\end{equation}
where we take into account the periodicity of Bloch-wave envelopes.
If one considers this relation only for a single point $\mathbf{r}$ in the unit cell, it becomes mathematically equivalent to the problems considered in spectral analysis of temporal series~\cite{Roy:1991-109:PRP, Mandelshtam:2001-159:RAR}, and high-resolution spectral methods can be used to extract the mode wavenumbers~\cite{Sukhorukov:2009-3716:OE}. However, the special property of periodic waveguides is that Eq.~\reqt{U} shall be satisfied simultaneously for all spatial locations $\mathbf{r}$ in the unit cell.
In order to determine the values of $k_m$ and $A_m(\mathbf{r};\omega)$ which describe most accurately the whole measured field, we employ the least squares method and seek a minimum of the functional
   $W = \int_\mathbf{r}\sum_{n=1}^N |w_n|^2\,\ud \mathbf{r} / \int_\mathbf{r}\sum_{n=1}^N |U_n|^2\,\ud \mathbf{r}$,
where integration is performed over the unit cell.
For given wavenumbers, the minimum $W_A(\{k_m\}) = {\rm min}_A W$ is achieved when $\partial W / \partial A_m = \partial W / \partial A_m^\ast = 0$. It follows that for each point  $\mathbf{r}$ in a unit cell, the optimal amplitudes satisfy the linear matrix equation $C^H \cdot C \cdot \widetilde{A}(\mathbf{r}) =  C^H \cdot \widetilde{U}(\mathbf{r})$, where components of vector $\widetilde{A}(\mathbf{r})$ are the optimal amplitude values, components of the matrix $C$ are $C_{np} = \exp( i k_p n )$, and vector $\widetilde{U}(\mathbf{r})$ components are
$U_n(\mathbf{r})$ for $p = 1:M$ and $n = 1:N$. 
We can show that $W_A(\{k_m\}) = W_{A=\widetilde{A}} = 1 - \int_\mathbf{r}\widetilde{U}^H(\mathbf{r}) \cdot C \cdot \widetilde{A}(\mathbf{r}) \,\ud\mathbf{r}/ \int_\mathbf{r} \widetilde{U}^H(\mathbf{r}) \cdot \widetilde{U}(\mathbf{r}) \,\ud\mathbf{r}$.
The remaining task is to find the absolute minimum $W_{\rm min} = \min_{k_m} W_A$ (note that, by definition, $W_A$ is real and positive), and this can be done numerically, for example by using the 'fminsearch' function in Matlab.

\pict{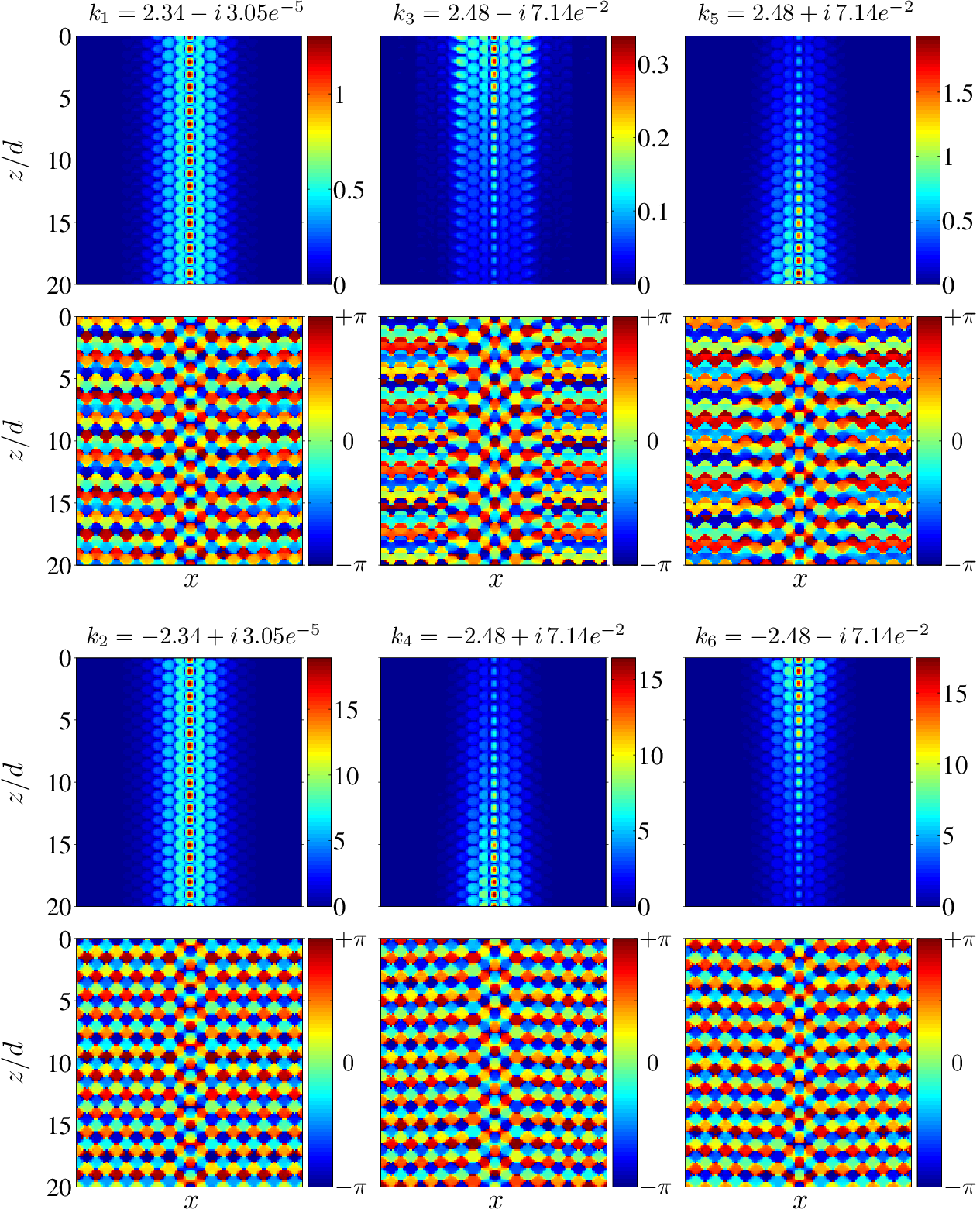}{profile}{Extracted magnitudes and phases of mode profiles $A_m(x,z)$ at the slow-light frequency based on the original electric field data shown in Fig.~\rpict{schematic}(b).
}

We apply our method to the two-dimensional [$\mathbf{r}=(x,z)$] photonic crystal waveguide shown in Fig.~\rpict{schematic}(a), where we indicate the unit cell with shading and illustrate the definition of $U_n$.
Photonic crystals are viewed as effective media for generating slow light because their dispersion characteristics can be controlled with design parameters~\cite{Vlasov:2005-65:NAT}. However, coupling light into a slow-light waveguide can be a key practical challenge due to the field mismatch between the incoming mode with high group velocity $v_g$ and the slow mode. The photonic crystal waveguide being analyzed here mediates efficient coupling into and out of a slow mode without any matching region, where an evanescent mode (PC 2) helps match the fields of the incoming mode (PC 1) and the slow mode (PC 2) without carrying any energy itself~\cite{White:2008-2644:OL}. We aim to extract the individual modes of PC 2 from a numerical data of $E(x,z)$, such as shown in Fig.~\rpict{schematic}(b). In PC 2, the dispersion relation contains an inflection point, and therefore the total number of modes that primarily define the light dynamics is $M=6$, accounting for the forward and backward slow-light ($m=1,2$) and evanescent modes ($m=3,4,5,6$). The fact that the waveguide has multiple modes and a relatively short length provides good testing environment for our method.

We note that the spatial spectral analysis can benefit through the application of additional constraints due to the symmetry of modes in lossless dielectric waveguides, wavenumbers of which are related as:
$k_2=-k_1$, $k_4=-k_3$, $k_5=k_3^\ast$, $k_6=-k_5$.
Therefore, $k_1$ and $k_3$ are the independent parameters that fully define the mode dispersion, and we take this into account for numerical minimization of the functional $W$.
The extracted wavenumbers $k_m$ are shown in Fig.~\rpict{wavenumber}(a) and they are in good agreement with numerical calculations. Most importantly, we simultaneously extract the profiles of the corresponding Bloch waves, which amplitudes reflect the excitation dynamics.
The mode profiles $A_m(x,z)$ at the slow-light frequency $d/\lambda\simeq0.2662$ are presented in Fig.~\rpict{profile}. The profiles clearly show the propagating ($m=1,2$) and evanescent ($m=3,4,5,6$) modes. This provides an essential insight into the light dynamics, which cannot be inferred directly from the field profiles as shown in Fig.~\rpict{schematic}(b), or from their SFT spectra.

To test our method in a possible experimental situation where noise is present, we add a normal distribution of random complex numbers to the electric field $U_n(x,z)$ with a mean of zero and a standard deviation of 0.05.
We have successfully recovered the mode dispersion under the effect of such perturbations.

In conclusion, we have presented a general approach for the simultaneous extraction of wavenumbers and amplitudes profiles of Bloch waves based on near-field measurements in periodic waveguides. We have demonstrated the application of this method for the simultaneous characterization of multiple propagating and evanescent modes, and its robustness under the effect of noise. This approach may serve to provide essential insight into the light dynamics in complex photonic-crystal circuits.

\end{document}